\documentclass[12pt,preprint]{aastex}    
\newcommand{\be}{\begin{equation}}  
\newcommand{\ee}{\end{equation}}    
\newcommand{\bea}{\begin{eqnarray}}
\newcommand{\eea}{\end{eqnarray}}

\def\BV{Brunt-V\"ais\"al\"a }

\slugcomment{}
\shorttitle{Simple Seismic Tests}
\shortauthors{Kennedy}

\begin{document}

\title{Simple Seismic Tests of the Solar Core\footnote{Contribution
to the INT Low-Energy Neutrino Physics Workshop, Univ. Washington, Seattle
(June-September 1999).}}
\author{Dallas C. Kennedy}
\authoraddr{Institute for Nuclear Theory, University of Washington\\ and\\
University of Florida}
\affil{Institute for Nuclear Theory, University of Washington, Box
351550, Seattle WA 98195\\ and\\
Department of Physics, University of Florida, Box 118440, Gainesville FL 32611}
\email{kennedy@phys.ufl.edu}
\affil{\rm DOE/ER/02A-40561-INT99-31 $\bf{\bullet}$ UF-IFT-HEP-99-7\\ 
Original: July 1999 $\bf{\bullet}$ Revised: February 2000}
\begin{abstract}
A model-independent reconstruction of mechanical profiles (density, pressure)
of the solar interior is outlined using the adiabatic sound speed and
bouyancy frequency profiles.  These can be inferred from helioseismology if
both $p$- and $g$-mode frequencies are measured.  A simulated reconstruction
is presented using a solar model bouyancy frequency and available sound
speed data.
\end{abstract}

\indent\centerline{\hskip-40pt To appear in {\bf the Astrophysical Journal}:
10 September 2000}

\indent Standard solar models (SSMs) are now very accurate, exemplified by 
the work of Bahcall et al. (Bahcall et al. 1998; Pinsonneault 
1998) and Turck-Chi\` eze et al. (Brun et al. 1998).
But confidence in predicted solar neutrino fluxes is enhanced by checks of the 
solar core depending on only a few general assumptions, independent of 
detailed models.  This can be done with helioseismology, although the 
$g$-modes are needed.  
Serious claims of $g$-modes observations have been put forward
(Hill \& Gu 1990; Thomson et al. 1995) but remain controversial
(Guenther \& Sills 1995; Lou 1996).  Nonetheless, the $g$-modes will
undoubtedly be observed at some point, despite
their attenuation by the convective zone.

We assume negligible radiation pressure $P_\gamma\ll P_{\rm gas}$
and no significant departures from an ideal gas $(P=P_{\rm gas}=
\rho\Re T/\mu )$.
\begin{eqnarray}
\Gamma = d\ln P/d\ln\rho\quad ,\quad \Gamma_{\rm ad} = (\partial\ln P/
\partial\rho )_{\rm ad}\quad ,\\ \nonumber
\nabla = d\ln T/d\ln P\quad ,\quad \nabla_{\rm ad} = (\partial\ln T/
\partial\ln P)_{\rm ad}\quad .
\end{eqnarray}
$\Gamma_{\rm ad}$ = 5/3, $\nabla_{\rm ad}$ = 2/5 for a monatomic gas.  The \BV 
(bouyancy) frequency $N(r):$
\begin{equation}
N^2 = \frac{g}{\lambda_P}\cdot\Bigl\lbrack\frac{1}{\Gamma}-
\frac{1}{\Gamma_{\rm ad}}\Bigr\rbrack = (\rho g^2/P)\cdot
\bigl\lbrack\nabla_{\rm ad}-\nabla +\nabla_\mu\bigr\rbrack\quad ,\quad
\lambda^{-1}_P = -d\ln P/dr\quad ,
\end{equation}
with $\nabla_\mu$ = $d\ln\mu /d\ln P$, is real for propagating $g$-waves,
outside any convective zones.  The base of the solar convective zone
$r_{\rm BCZ} = (0.709)R_\odot$ (Gough et al. 1996).
The adiabatic squared sound speed $c^2_{\rm ad}$ = $\Gamma_{\rm ad}\cdot 
P/\rho$.  

The adiabatic seismology formalism here follows Unno et al. (1989), with some 
differences in notation.  
The seismic modes are labeled by radial index $n$ (number of radial
nodes) and angular indices $l$ and $m$; frequencies degenerate in the last
if rotation is ignored.  The spectrum exhibits ``middle'' $f$ (fundamental or 
$n$ = 0, $l >$ 1) modes and the $p$- $(g$-) modes rising above (falling below)
the $f$ mode in $\nu$ for $n >$ 0, $l >$ 0.
For large $n$, the $p$- and $g$-modes are concentrated near the 
surface and the center, respectively.  Their frequencies $\nu_p$ and $\nu_g$ 
are
\begin{equation}
\nu^{-1}_p(n,l)\simeq \frac{2}{n+\frac{1}{2}\sqrt{l(l+1)}}
\int^{R_\odot}_0\frac{dr}{c_{\rm ad}(r)}\quad ,\quad
\nu_g(n,l)\simeq \frac{\sqrt{l(l+1)}}{2n\pi^2}\int^{r_{\rm BCZ}}_0 
\frac{dr}{r}\ N(r)\quad ,
\end{equation}
for large $n$.  $(l$ cannot be large in $\nu_p.)$  A large set of measured 
frequencies allow reconstruction of the $c_{\rm ad}(r)$ and $N(r)$ profiles.
The Bahcall-Pinsonneault 1998 (BP98) SSM is used here when specific numerical 
results are needed.

\section{Tests of Hydrostatic Equilibrium}

Present helioseismic data rule out the $\Gamma$ = 5/3 convective polytrope 
deep into the core (Basu et al. 1997; Christensen-Dalsgaard 1997).
Hydrostatic equilibrium is tested by correct prediction of the $\nu_g(n,l)$, 
$n\gg$ 1.  Success here would limit any remaining
possibility of convection at the center (otherwise $N^2 < 0)$, although the Sun 
is not far from this state $(M\simeq 1.1 M_\odot$ in standard calculations;
see Kippenhahn \& Weigert 1990).  An equivalent test of hydrostatic equilibrium at
the center is provided by the sound speed slope, measurable in principle with
$p$-modes alone:
\begin{equation}
{dc^2_{\rm ad}\over{dr}} = -g\Gamma_{\rm ad}\cdot\Bigl( 1-\frac{1}{\Gamma}
\Bigr) + \frac{P}{\rho}\cdot\frac{d\Gamma_{\rm ad}}{dr}\quad ,\quad
1-\frac{1}{\Gamma} = \nabla -\nabla_\mu\quad .
\end{equation}
Assume $d\Gamma_{\rm ad}/dr$ = 0 in the core and a non-singular mass distribution,
so that
\begin{equation}
g = Gm/r^2 = (4\pi G/3)\rho_cr\rightarrow 0~{\rm as}~r\rightarrow 0\quad .
\end{equation}
Thus $dc^2_{\rm ad}(0)/dr$ must vanish.

The measured sound speed slope places a limit on an unperturbed
state with non-zero velocity field ${\bf v}_0$, taken here as radial for
simplicity.  This test checks for the presence of convective motion that does
not efficiently transfer heat.  Such convection would not directly affect $\Gamma$
and the $P/\rho$ profile and thus is not ruled out in general by current $p$-mode 
helioseismic data.  Hydrostatic equilibrium is modified to
\begin{equation}
-\frac{dP}{dr} = \rho g +\rho v_0\frac{dv_0}{dr}\quad ,
\end{equation}
ignoring spherical-symmetry-breaking rotation and magnetic effects, also 
calculable.  Assume again $d\Gamma_{\rm ad}/dr$ = 0 in the core and $g(0)$ = 0.
Equation~(4) no longer vanishes at the center and instead yields
\begin{eqnarray}
\frac{dc^2_{\rm ad}(0)}{dr} = -\Gamma_{\rm ad}\cdot v_0(0)\frac{dv_0(0)}{dr}\cdot 
\Bigl[1-\frac{1}{\Gamma (0)}\Bigr]\quad .\nonumber
\end{eqnarray}
Equivalently, in terms of $g$-modes, $\nu_g(n,l)$ is corrected 
(to linear order in $v_0)$
by the ``expectation value'' in the $(n,l)$ mode of $k_rv_0/2\pi$, where $k_r$ 
= radial wavenumber.  The related length scale $\approx R_\odot$/few for 
low-$n$ modes; but for high-$n$ $g$-modes concentrated near the center, it is 
the central 
scale $R_0:$ $R^2_0$ = $3P_c/2\pi G\rho^2_c$, with $R_0/R_\odot$ = 0.12 for the 
present SSM (Kennedy \& Bludman 1999).  For large $n$,
\begin{equation}
k_r\rightarrow \frac{n}{R_0\Omega_g}\cdot\sqrt{{8\pi^3\rho_cG\over 3}
\Biggl[\frac{1}{\Gamma (0)} - \frac{1}{\Gamma_{\rm ad}}\Biggr]}\quad ,\quad
\Omega_g = \int^{r_{\rm BCZ}}_0\frac{dr}{r}\ N(r)\quad .
\end{equation}
A conservative relative observational error for $\nu_g$ is 0.003, while the relative
theoretical uncertainty $\simeq$ 0.001 (Hill \& Gu 1990; Guenther \& Sills 1995;
Bahcall et al. 1998).  Thus a direct bound on the
core $v_0\lesssim$ $2\pi\nu_g(n,l)/k_r\sim$ $(7.5\times 10^7 {\rm \ cm\ 
sec}^{-1})\sqrt{l(l+1)}/n^2$, or 1.2$\times 10^5$ cm sec$^{-1}$ for $l$ = 1, 
$n$ = 30, is feasible, independent of azimuthal splitting by $m$.
 
\section{Mechanical Profiles}

The $N(r)$ and $c_{\rm ad}(r)$ profiles are sufficient to reconstruct uniquely 
the mechanical (hydrostatic) profiles.
$N(r)$ and $dc_{\rm ad}(r)/dr$ are proportional, either one controlling the 
$\nu_g(n,l):$
\begin{equation}
\frac{dc_{\rm ad}(r)}{dr} = -\frac{N(r)/2}
{\sqrt{\Gamma_{\rm ad}/\Gamma (r) -1}}\cdot
\Bigl\lbrack 1-\frac{1}{\Gamma (r)}\Bigr\rbrack\Gamma_{\rm ad}\quad .
\end{equation}
Gravity modes are a result of the inhomogeneity of $c_{\rm ad}(r)$.
Assuming $d\Gamma_{\rm ad}/dr$ = 0, $\Gamma$ satisfies a quadratic equation,
\begin{eqnarray}
(1+{\cal D})\Gamma^2 - (2+{\cal D}\cdot\Gamma_{\rm ad})\Gamma +1 = 0\quad ,
\quad {\cal D}(r)\equiv \Biggl\lbrack{2\cdot dc_{\rm ad}(r)/dr\over{\Gamma_{\rm ad}
\cdot N(r)}}\Biggr\rbrack^2\quad ,
\end{eqnarray}
with one physical root,
\begin{equation}
\Gamma (r) = {2+{\cal D}(r)\cdot\Gamma_{\rm ad}+
\sqrt{[2+{\cal D}(r)\cdot\Gamma_{\rm ad}]^2-4[1+{\cal D}(r)]}\over{2[1+
{\cal D}(r)]}}\quad .
\end{equation}
Thus the $\Gamma (r)$ profile is reconstructible with $N(r)$ and a numerical
derivative of $c_{\rm ad}(r)$.  Note that $\Gamma$ = 1 when $dc_{\rm ad}/dr$
and ${\cal D}$ = 0.

The mass and pressure profiles then follow:
\begin{eqnarray}
Gm(r) & = & -\frac{r^2\cdot dc^2_{\rm ad}(r)/dr}{\Gamma_{\rm ad}\cdot [1-
1/\Gamma (r)]}\quad ,\\ \nonumber
4\pi G\rho_c & = & -\frac{6\cdot dc^2_{\rm ad}(0)/dr^2}{\Gamma_{\rm ad}\cdot
[1-1/\Gamma (0)]}\quad ,\\ \nonumber
P(r) & = & \frac{\rho (r)\cdot c^2_{\rm ad}(r)}{\Gamma_{\rm ad}}\quad ,
\end{eqnarray}
where $\rho (r)$ is obtained from $m(r)$.  Note that $m>0$, because
$\Gamma >1$ if $dc^2_{\rm ad}/dr<0$ and $\Gamma <1$ if $dc^2_{\rm ad}/dr>0$.
(The $dc^2_{\rm ad}/dr=0$ and $r\rightarrow 0$ limits are well-defined in
the ratios.)  The present SSM inner core is an example of the $\Gamma <$ 1 
case.  Dimensionless structure in terms of homology variables can also be 
derived (Kennedy \& Bludman 1999).

Direct methods of inversion without a reference SSM, based on $p$-modes alone
and including use of the sound speed slope as a diagnostic of the equation of
state, are discussed in the older literature (Gough 1984; Christensen-Dalsgaard
et al. 1985; Gough \& Toomre 1991; Gough et al. 1995) and, since the advent 
of SOHO, GONG, and BiSON/LOWL, by Gough et al. (1996), Basu et al. (1997), 
and Christensen-Dalsgaard (1997).  Since the $\nu$, especially at long 
wavelength, are affected by the whole Sun, errors in the inferred spatial
structure are strongly correlated.  An inversion for $N(r)$, like the 
inversion for $c_{\rm ad}(r)$, will have uncertainties in both $N$ and the
position $r$.  The derived profiles~(11) will have errors compounded from
all these sources: $c_{\rm ad}(r)$, $N(r)$, and $r$, and thus larger than 
errors in any one.

\section{Numerical Test of the Method}

As the SSM evolves, $\nabla$ falls while $\nabla_\mu$ increases from zero, 
the core becoming more concentrated.  Helioseismology is adiabatic and 
non-evolutionary and cannot give the thermal or chemical structure without 
further assumptions or a full SSM.  Many SSM core features can
be understood via mechanical-thermal core homology without a detailed model
(Bludman \& Kennedy 1996).  
Homology applied to the {\it core only} is valid for power-law opacity and 
luminosity generation and does {\it not} require a polytrope.  A polytrope 
is not even approximately valid for the present SSM core in any case 
(Kennedy \& Bludman 1999).  The homology in reality is violated somewhat, 
as the exponents are not constant and the luminosity is not exclusively 
produced by one reaction chain.\footnote{If the specific luminosity production 
is $\varepsilon$ = $\varepsilon_o\rho^\lambda T^\theta$ and the specific 
opacity is $\kappa$ = $\kappa_o\rho^qT^{-s}$, the homology yields 
$r^{\theta -s+3\lambda +3q}\sim\varepsilon_o\kappa_o\cdot\mu^{\theta -s-4}
\cdot m^{\theta -s+\lambda+q-2}$ (correcting an error in Bludman \& Kennedy
1996).}  But an accurate thermal and chemical reconstruction is not much 
less complicated and requires no fewer assumptions than a full SSM.  On the
other hand, the mechanical structure of a SSM can be specified by the 
model-independent reconstruction of Section~2, rather than calculated.

As $\rho$ and $P$ are in principle inferable from helioseismology, the
numerical precision of the reconstruction can be estimated.  No $g$-mode
measurements are now available, so the SSM bouyancy frequency is used
here.  Combining $c_{\rm ad}(r)$, $dc_{\rm ad}(r)/dr$, and the SSM 
$N(r)$ and assuming $\Gamma_{\rm ad}$ = 5/3, the profile ${\cal D}(r)$ is
obtained~(9) and hence $\Gamma (r)$.

Figure~1 shows the measured sound speed below the convection zone, the points
being fixed by helioseismic inversion (Christensen-Dalsgaard 1997).  The BP98 
SSM $\Gamma$ and $N$ profiles, respectively, are shown in Figs.~2(a,b).  In 
Figure~3 is displayed the ratio of the reconstructed to the SSM $\Gamma (r)$.  
The helioseismic inversion points are separated by steps of 
$\Delta(r/R_\odot )\simeq$ 0.01, and we should thus expect agreement with
the SSM at level of about one percent, consistent with Fig.~3.
The inner boundary at $r/R_\odot$ = 0.05 marks the terminus of the present 
helioseismic inversion, the outer boundary the base of the convective zone at 
$r_{\rm BCZ}/R_\odot$ = 0.71.  Cutting off the fit at both ends introduces 
artificial errors, and thus the first three and last three points are not 
shown in Fig.~3.

The agreement of a true reconstruction with the SSM should be better than
Fig.~3, but the overall uncertainty would be somewhat larger than a percent, 
reflecting the $c_{\rm ad}(r)$ uncertainty; the additional uncertainties due 
to a measured, rather than a model, $N(r)$ profile; and the errors induced
by spatial correlation of the helioseismic inversion mesh points.

\section{Summary}

Some solar core tests are proposed, independent of model details.  Measured
$p$- and $g$-modes are needed to reconstruct $c^2_{\rm ad}(r)$ and $N^2(r)$
uniquely and independently of a reference SSM, with only general physical 
assumptions: hydrostatic equilibrium, the ideal gas law, and known, constant
$\Gamma_{\rm ad}$.

Absence of core convection implies $N^2 >$ 0 in~(2) and the existence of 
$\nu_g(n,l)$ for high $n$~(3).  A non-singular, static mass distribution 
implies vanishing of the sound speed slope~(4) at the center~(5).  The
measured slope and/or high-$n$ $\nu_g$ can set a limit $\sim{\cal O}
(10^5)$ cm sec$^{-1}$ on a central velocity field.  Reconstruction of the solar
interior mechanical structure follows~(10-11) from $c_{\rm ad}(r)$,
$dc_{\rm ad}(r)/dr$, and $N(r)$.  A test of the method is presented, 
using the available, measured $c_{\rm ad}(r)$ and a model $N(r)$ 
profile.\footnote{
Numerical analysis code in Mathematica 3.0 with data and model tables
can be downloaded at http://www.phys.ufl.edu/\~{}kennedy/soft/cover.html.}

The $c_{\rm ad}$ and $N$ profiles to 
the origin are essential for testing the SSM in the inner core, the critical
region for the solar neutrino problem, where the higher-energy Be and B 
neutrinos are produced and where solar structure becomes strongly 
non-polytropic (Bahcall 1988; Kennedy \& Bludman 1999).  $p$-mode data on
a finer radial mesh and extending to the center are presently lacking, but
might become available soon, even if $g$-mode measurements remain elusive.

Truly independent tests of thermal
structure and chemical evolution would require comparison with other Sun-like
stars, in particular via asteroseismology (Deubner et al. 1998).  With the
appropriate changes, the method outlined here can be used to analyze such 
stars as well.

\acknowledgments

The author thanks John Bahcall (Institute for Advanced Study) and Marc 
Pinsonneault of (Ohio State Univ.) for the BP98 model tables and J\o rgen 
Christensen-Dalsgaard (Aarhus Univ.) for helioseismic data.\footnote{
For Bahcall-Pinsonneault present SSM, see 
http://www.sns.ias.edu/\~{}jnb/SNdata/solarmodels.html.}
This work is partly based on research done in collaboration with Sidney 
Bludman (Univ. Pennsylvania and DESY) and
was supported by the Department of Energy under Grant Nos.
DE-FG06-90ER40561 at the Institute for Nuclear Theory (Univ. Washington) and
DE-FG02-97ER41029 at the Institute for Fundamental Theory (Univ. Florida),
and by the Eppley Foundation for Research.  The author is also grateful to the 
Aspen Center for Physics for its hospitality.

\newpage
\centerline{\bf FIGURES}

\figcaption{
The $p$-mode adiabatic sound speed in core of Sun inferred by helioseismic
inversion.  Uncertainties too small to show.  (Data courtesy of 
J.~Christensen-Dalsgaard.)\label{FIG1}}

\figcaption{
The BP98 present Sun model profiles from center to base of convective zone 
at $r/R_\odot$ = 0.71: (a) stiffness $\Gamma$, with $\Gamma\rightarrow$ 5/3 
at base of convective zone; (b) \BV bouyancy frequency $N$, which vanishes 
at center and base of convective zone.  (Model tables courtesy of 
M.~H.~Pinsonneault.)\label{FIG2}}

\figcaption{
Profile of ratio of simulated reconstruction of $\Gamma$ to BP98 SSM $\Gamma$,
extending from $r/R_\odot$ = 0.075 to 0.68.\label{FIG3}}

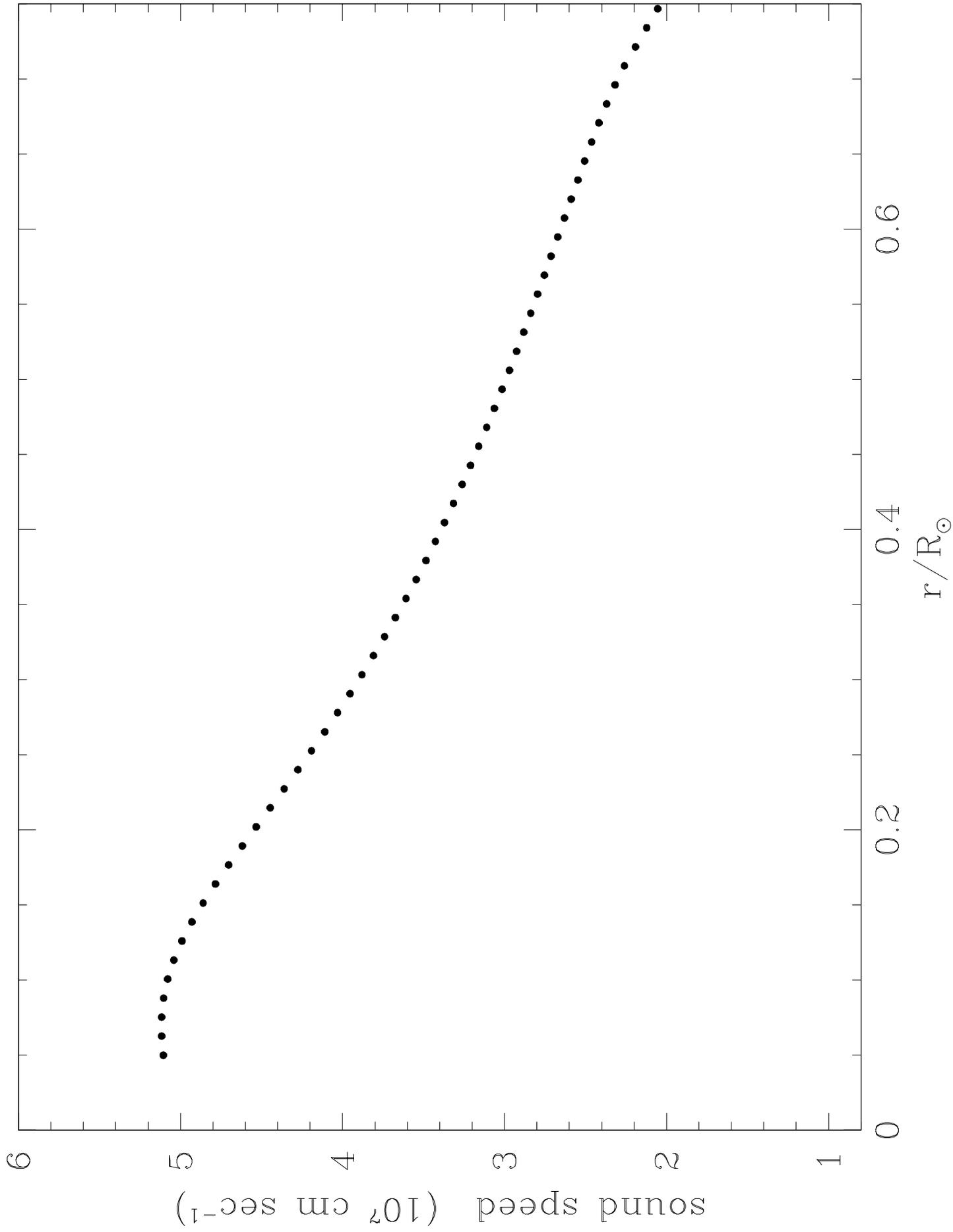
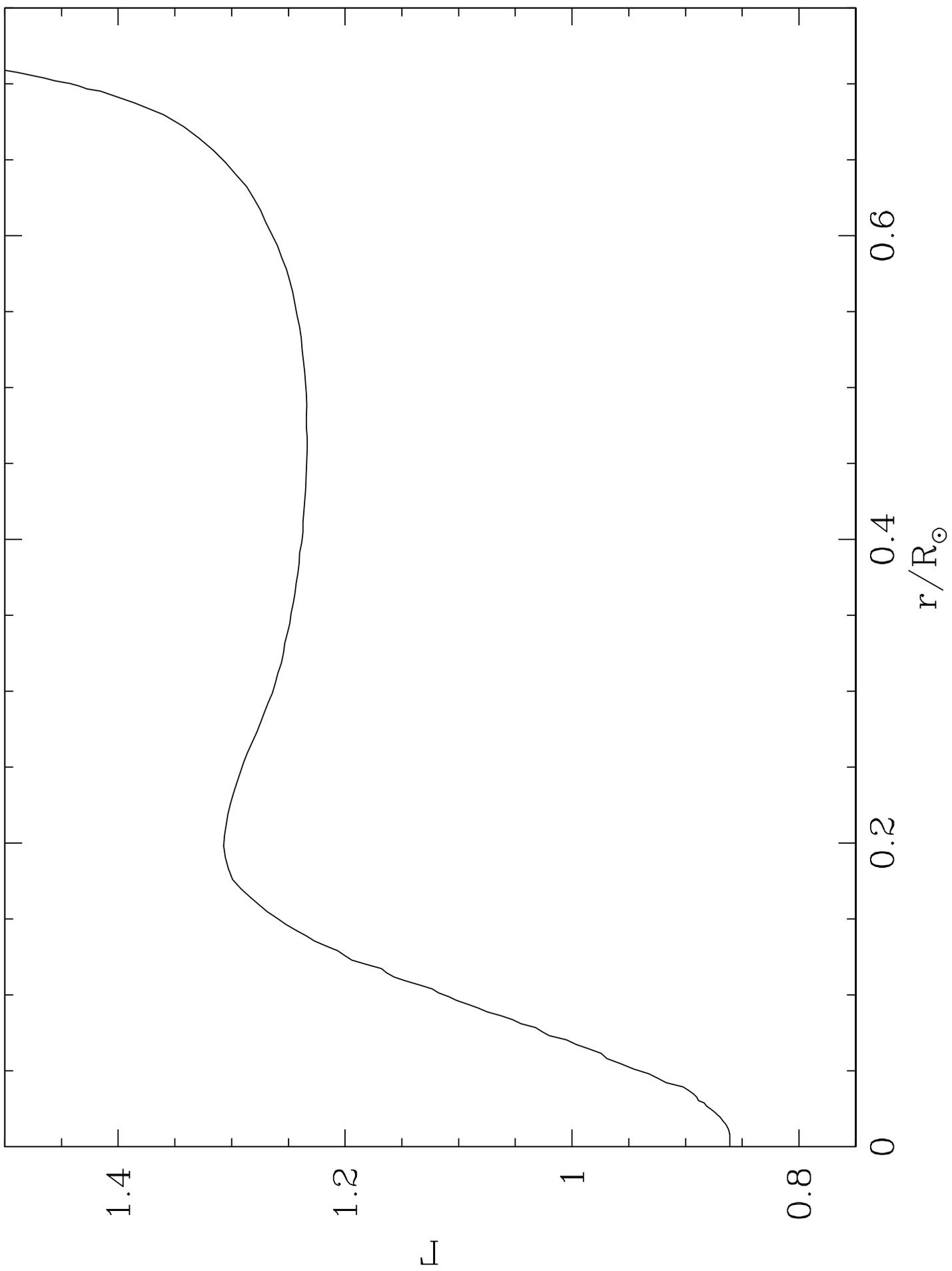
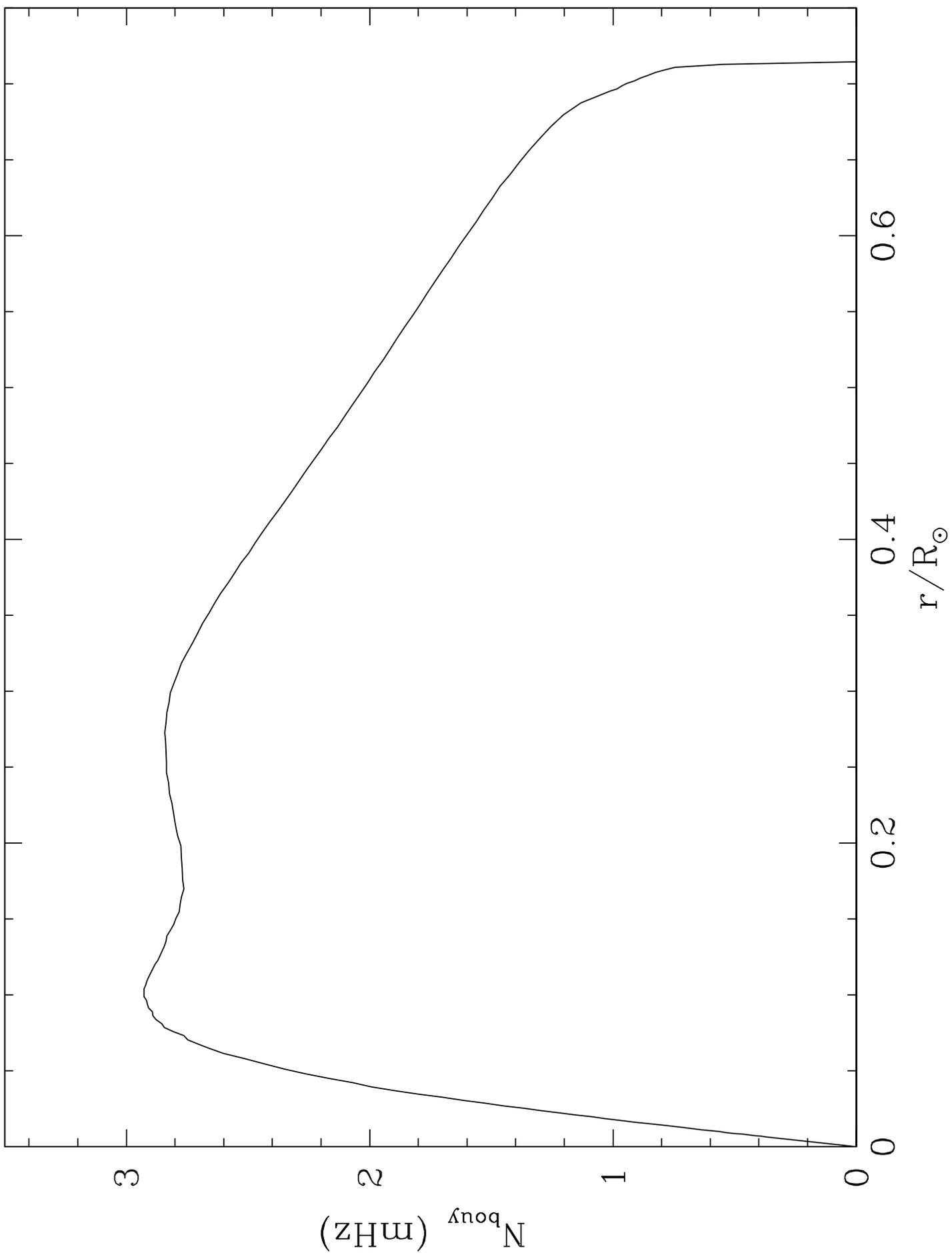
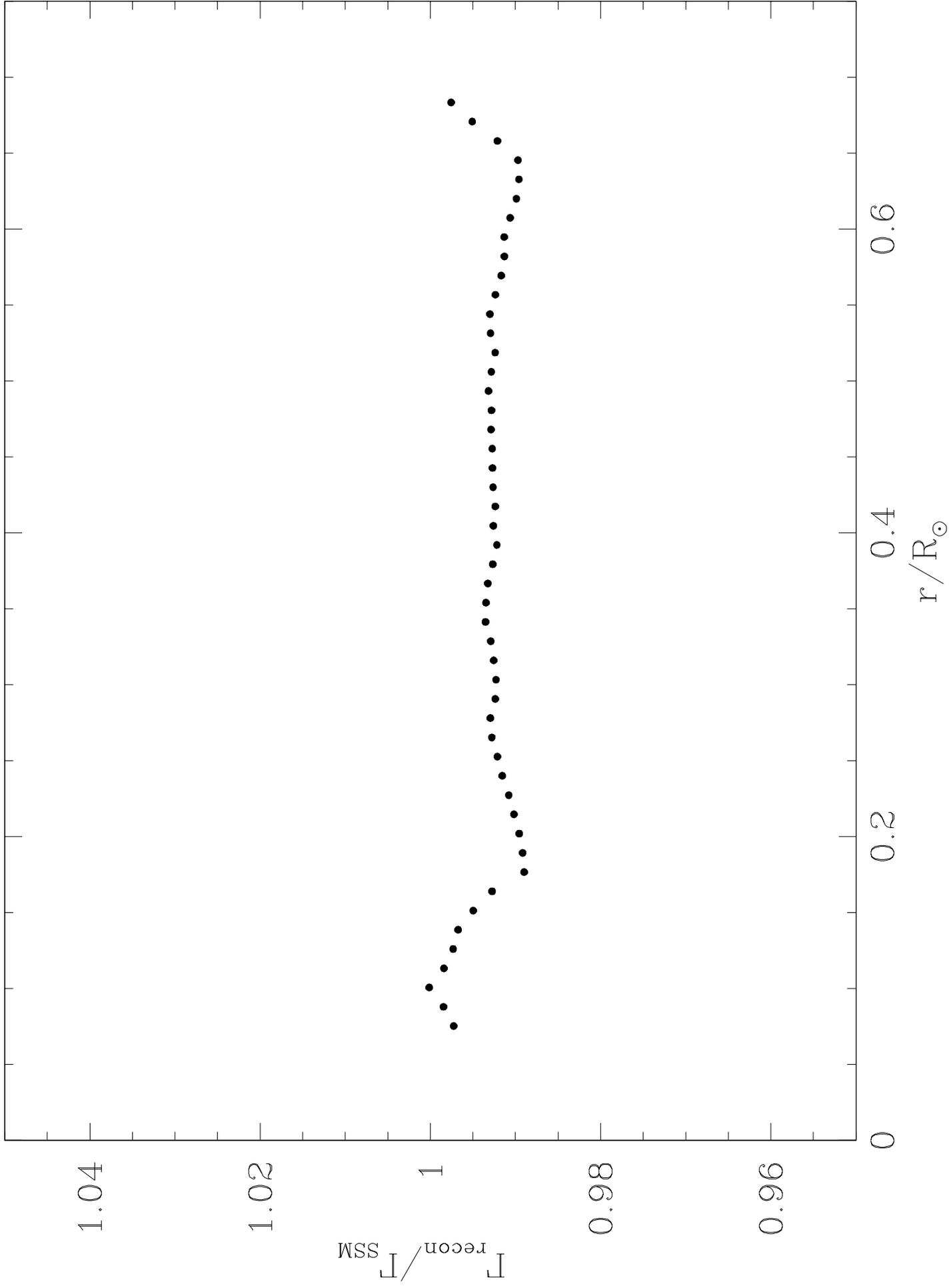
\end{document}